# Quartier d'artistes versus cluster numérique. Entre conflit foncier et production d'un nouvel espace créatif : le 22@ de Poblenou à Barcelone

**Patrice BALLESTER** Docteur en géographie, aménagement, urbanisme, paysage urbain
Professeur d'histoire-géographie
Rectorat de l'Académie de Toulouse,
Place Saint-Jacques
BP 7203
31073 Toulouse CEDEX 7
patrice.ballester@ac-toulouse.fr

**Résumé**
Poblenou (ou le nouveau village, en catalan) est un quartier de Barcelone connaissant depuis plus de dix ans une régénération urbaine basée sur le concept de la nouvelle économie des TIC (Technologies de l'Information et de la Communication) comme source d'une nouvelle image attractive pour la localisation d'entreprises internationales dans le cadre d'un projet de district créatif nommé 22@. Une transition territoriale produit une gentrification contrariée par les artistes, les intellectuels et la population locale qui ont adopté comme emblème de leur action le sauvetage d'un fleuron de l'architecture industrielle : Can Ricart, lieu de résidence de nombreux artistes. Depuis, la planification urbaine a été repensée dans ses marges et le paysage urbain du quartier tend à refléter le consensus entre politiques, firmes multinationales et artistes rescapés de cette forte pression immobilière. Ces dits artistes deviennent des référents, tout en essayant de s'intégrer dans la thématique de la nouvelle économie créative du territoire par des alliances, des commandes et des coopérations timides mais régulières avec le monde marchand. La nouvelle Soho européenne recherche encore une identité ainsi qu'un certain art de vivre fondé sur les nouveaux rapports de force mis en valeur par les artistes dans leurs critiques, leurs actions et leur communication sur :
• la cohabitation entre anciens et nouveaux habitants ;
• le choix entre architecture contemporaine et protection patrimoniale ;
• l'artiste comme créateur, assistant ou complice de la classe dite créative.

*Mots clefs* : Barcelone, artistes, cluster créatif, conflit, identité

**Abstract**
Poblenou (or the new village in Catalan) is a district of Barcelona known for over ten years urban regeneration based on the concept of the new economics of ICT (Information and Communications Technology) as a source of new attractive image for the location of international companies with a creative project called 22@ district. A transition produced a territorial gentrification thwarted by artists, intellectuals and local people adopting as the emblem of their action, the rescue of a flagship industrial architecture: Can Ricart, home of many artists. Since, urban planning has been redesigned in its margins and townscape of the area tends to reflect the consensus between politicians, multinational companies and artists survivors of this high-pressure real estate. These artists became referents, while trying to fit into the theme of the new creative economy of the territory through alliances, orders and cooperation shy but regular. The new european Soho research yet an identity and a certain lifestyle based on the new balance of power highlighted by the artists in their criticism, actions and communications:
•       cohabitation between old and new residents;
•       the choice between contemporary architecture and heritage protection;
•       the artist as creator, assistant or complicit in the so-called creative class.

**INTRODUCTION**

Poblenou, ou le nouveau village (en catalan), est un quartier de la commune de Barcelone dont la physionomie change depuis plus d'une dizaine d'années. Un nouveau paysage urbain comprenant de grandes tours tertiaires issues des ambitions architecturales de sociétés internationales se découvre aux habitants de cette zone historique et symbolique de la capitale catalane. La spéculation immobilière y est forte et repose sur la confrontation entre deux visions territoriales de cet espace péri-central d'une métropole européenne dynamique : pour les uns, un territoire d'artistes, de grandes friches industrielles à sauvegarder et des logements bon marché ; pour les autres, un territoire hautement spéculatif et stratégique de la nouvelle économie préfigurant de part sa situation de crise urbaine un futur espace de la mondialisation (Scott, 1999). Cette dichotomie montre les représentations ambivalentes des multiples acteurs qui interviennent sur ce paysage contemporain en transition, composé de vastes friches délaissées, de noyaux d'habitation anciens, d'une *Rambla* désuète, d'usines désaffectées et, récemment, de gratte-ciels parfois plus œuvres d'art que lieux du néo-tertiaire (Ballester, 2013 b).

L'ensemble s'apparente à un surgissement de la ville et à un questionnement de son urbanité renouvelée. Que faire de cet espace emblématique, moteur de la Révolution Industrielle pour l'Espagne, comme le rappellent les nombreux articles de presse catalans (Andreu, Gol, Recio, 2001) au tournant des années 2000 ? Une régénération urbaine imposée doit accompagner et modifier le projet urbain barcelonais avec une ambition économique basée sur l'accueil des entreprises créatives reposant sur cinq thématiques : le design, la création graphique, le multimédia et les TIC appliquées au médical[1]. Lutter contre les délocalisations et proposer un nouveau modèle de vie/ville/village urbain pour attirer des entreprises de pointe devient un nouveau paradigme d'action, d'après les élites économiques catalanes qualifiant leur opération de projet/processus à caractère cumulatif (Clos, 2010 : 78). Mais ce modèle peut être considéré comme une énième utopie spatiale, un prototype dont le but est la fondation d'une nouvelle communauté reposant sur une mixité sociale souvent illusoire et s'immisçant progressivement dans la gouvernance urbaine, pour faciliter la sérendipité (Perrin, Soulard, 2010 : 74) ou affirmer la création d'une ville durable.

Les temporalités de la mutation de cette friche urbaine et la réalité du rôle des artistes dans ces transformations sont à analyser au regard de la constitution d'un rapport de force public-privé, déséquilibré et fluctuant (Ambrosino, Andres, 2008). Les artistes barcelonais, en association avec la population locale, sont-ils des spectateurs d'une mutation économique et sociale d'un territoire qu'ils ne peuvent empêcher de basculer dans le « tout économique », ou bien sont-ils des acteurs rappelant les bases de la vie en société et infléchissant le juridisme en cours du projet de cluster ? Le rôle des artistes dans la dynamique économique des territoires créatifs dans une grande métropole méditerranéenne est à questionner, voire à pondérer. Dans cette étude, nous prendrons position par rapport aux travaux d'Elsa Vivant, qui entrevoit une *« instrumentalisation des lieux culturels off dans les stratégies urbaines »*, avec des artistes qui sont *« plus des témoins ou des indicateurs de la gentrification que des déclencheurs ou des catalyseurs »* (Vivant, 2006b : 360 - 370 ; Vivant et Charmes 2008 : 60-61). Nous apporterons des nuances dans les temporalités de transformations urbaines et dans la capacité de résilience des artistes de Poblenou qui utilisent justement l'adjectif *créatif* pour proposer des contre-projets. Certes, le terme de « classe créative » est un marqueur urbain à l'identification imprécise (Vivant, 2006a ; Paquot, 2010 : 71), mais plutôt que de rester sur cette dernière notion controversée de Richard Florida, il est préférable de s'attarder avant tout sur la recherche de la qualité du service urbain proposé, dans une logique de l'offre qui l'emporte ou se superpose à celle de la demande, et donc qui implique la distinction, l'originalité et l'efficacité métropolitaine indispensable au dynamisme des villes mondiales, « *the power of place and quality of place* » (Florida, 2002 : 231-234). Le succès, critiqué, dudit « modèle » de cité méditerranéenne (Montaner, 2006 ; Capel, 2005) donne à certains agents municipaux l'idée de passer à une autre échelle d'intervention que celle des aires de nouvelles centralités des années 1980 (Acebillo, 1988). Comme nouveau cœur de ville, on travaille à l'échelle d'un district de plus de 200 hectares, Poblenou, rebaptisé 22@.

---

[1] Pour cette étude, nous analysons les principaux rapports d'état d'exécution des travaux du quartier de Poblenou entre 1999 et 2010, notamment les plans spéciaux des infrastructures et de révision du POS de l'an 2000 ainsi que le plan de protection du patrimoine architectural de 2006 consultable aux archives du siège de l'agence municipale : 22 Arroba BCN, S.A., Carrer d'Àvila, 138, 08018 Barcelona. Site internet : www.22barcelona.com au 01/11/2011.

Notre démarche se structure en trois interrogations reposant à la fois sur la théorie, la pratique et l'observation d'un terrain d'étude parcouru à plusieurs reprises entre les années 2004 et 2011.

*Première interrogation* : la ville créative et le rôle des artistes ne s'enracinent-ils pas dans un contexte socio-historique, trop souvent oublié, qui structure et explique en grande partie les évolutions futures et les convulsions dans les domaines de l'aménagement de l'espace et de son fonctionnement soi-disant autonome pour la création d'une société innovante, durable et ouverte ? *Deuxième interrogation* : la logique de montée en puissance, par une gestion foncière complexe, d'un cluster dit créatif dans le péri-centre barcelonais est-elle en contradiction ou en adéquation avec la logique des artistes de Poblenou, leurs activités et leur emprise au sol ? *Troisième interrogation* : peut-on évaluer les tentatives de collaboration entre les acteurs du 22@, notamment les entreprises, et faire un bilan à mi-chemin du projet, sans toutefois omettre la gravité de la situation économique espagnole ? L'ensemble de ces analyses nous permettra de dresser un bilan des images positives et négatives résultant de l'entreprise 22@. Dans cette optique, les conditions d'appropriation d'un espace symbolique se doivent d'être mises en perspective (1) pour justement comprendre les caractéristiques d'une production d'un espace de compromis émanant de multiples visions et actions des acteurs sur un territoire en transition forcée (2).

## 1. LES CONDITIONS D'APPROPRIATION D'UN ESPACE SYMBOLIQUE : POBLENOU

Pour des raisons historiques, géographiques et culturelles d'appropriation et d'aménagement de l'espace urbain en Espagne et Catalogne, le quartier de Poblenou devient un enjeu majeur pour la nouvelle Barcelone du XXIe siècle (1.1). La situation de crise du quartier est un élément fondateur de la confrontation de deux logiques d'appropriation de l'espace (1.2). De fait, il faut comprendre un combat symbolique cristallisant les mécontentements des groupes en présence et débouchant sur une solution de compromis à partir de la lutte pour la conservation de l'usine Can Ricart (1.3).

### 1.1. Le barri de Poblenou, un espace urbain doublement stratégique : mémoire et géographie urbaine

Photo 1 : Le quartier de Poblenou en 1992, entre rocade littorale et Diagonal de Cerdá

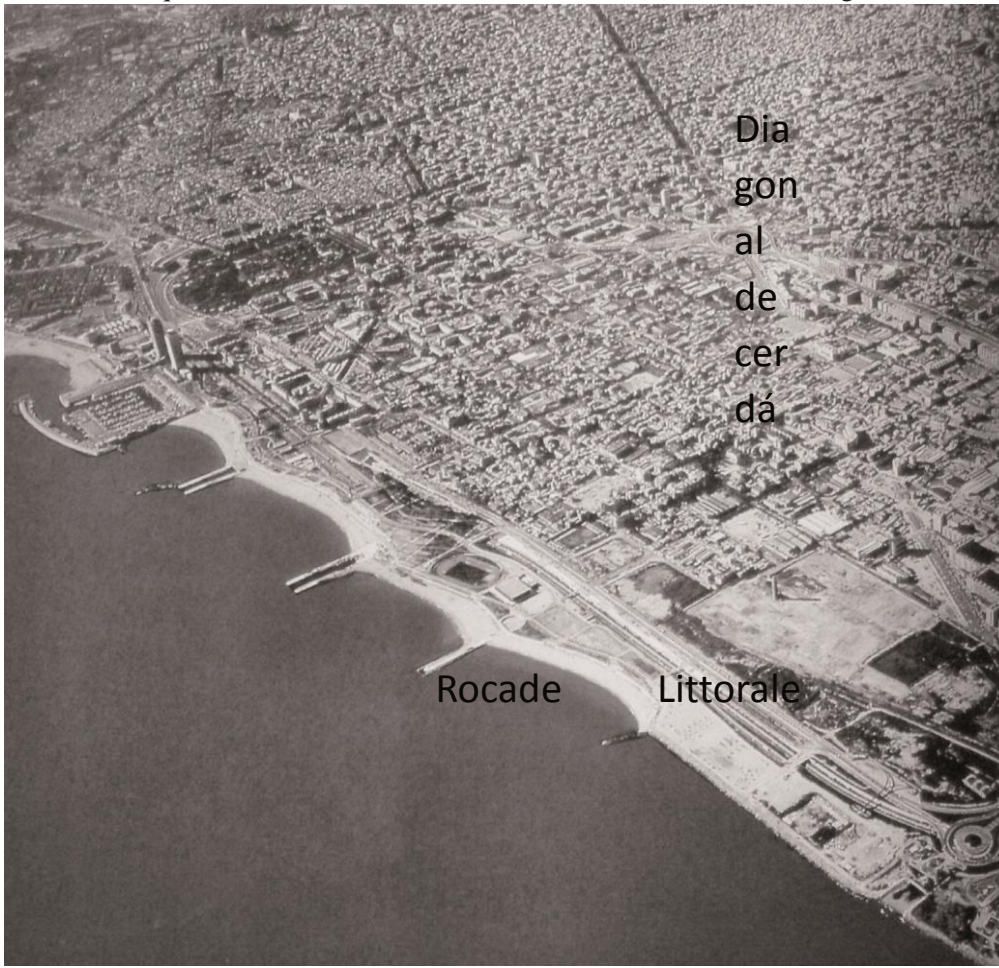

Source : Mairie de Barcelone, archives « division plan et photographie »        .

Les artistes sont-ils les relais d'une mémoire collective des conflits sociaux ? Les possibilités foncières offertes par de vastes friches et la situation stratégique du quartier font de Poblenou un quartier recherché par les artistes, dans un premier temps, et par les promoteurs, les PME et les firmes multinationales (FMN), dans un second temps. Une géohistoire du quartier apporte des éléments de réponse à la compréhension de cette renaissance urbaine. Avec la Révolution industrielle, la fabrique d'indiennes (tissus) débute par la réalisation de teinture et la présence de forges, comme celle d'Achon i Ricart. Point important concernant la gestion urbanistique de Poblenou : Cerdá décide, dans son plan d'*ensanche* de 1859 (extension urbaine planifiée en Espagne), la création d'une communauté : la Nouvelle Icare (*Nova Icaria*), sur le modèle utopique socialiste français du phalanstère (Coudroy de Lille, 1994). Le quartier de San Martí/Poblenou concentre la majeure partie des connexions ferroviaires de la ville avec la première ligne de chemin de fer espagnole Mataró-Barcelone inaugurée en 1848 pour le transport de marchandises et de charbon. On note la présence des institutions de crédit, des boutiques, des fabricants, des usines, des fonctionnaires douaniers et de sociétés internationales comme Martini et Rossi spiritueux (Checa, 2003). Il en ressort une impression générale d'un lieu où se créent des formes, des objets, des vêtements ainsi que des slogans et une identité urbaine d'innovation industrielle. Escofert est la plus grande fabrique de faïence d'Espagne au XIXe siècle. La question du foncier et des activités artistiques, culturelles et sportives est déjà objet de polémique tant pour l'implantation d'usines industrielles polluantes qu'à propos de la part à réserver aux logements d'ouvriers ou aux centres de loisirs et de récréation (Huertas, 2001). Les années 1960 permettent à la population du quartier de s'organiser contre des plans de promoteurs immobiliers sur la zone littorale du quartier. Le Plan de la Ribera en 1964/1965 puis le plan Porcioles au début des années 1970 proposent, avec l'aide d'une grande banque et de grandes entreprises, d'expulser une partie des habitants au profit de complexes immobiliers de rapport pour attirer une population aisée en lien avec le tourisme et l'interface maritime[2]. Le dernier plan implique l'expulsion de plus de 15 000 habitants, une perte d'emplois du fait de la démolition de nombreux ateliers, et des effets négatifs sur le voisinage à cause de la hauteur des immeubles projetés (Busquets, 2005 : 330-334). C'est un point de cristallisation et de mécontentement de toute la société barcelonaise qui voit dans ces projets la main des protégés du régime franquiste. S'ensuivent des manifestations, des pamphlets et des débats publics dans une presse catalane s'autorisant, dans les derniers temps de la dictature, une critique du régime et de ses ambitions immobilières (Tatjer, 1973). Cinq ans plus tard, le projet est abandonné en même temps que revient la démocratie en Espagne et la volonté de proclamer un « droit à l'urbanité ». Cette expression reflète la demande de la société civile catalane concernant de nouveaux espaces de dialogue et de participation à la vie publique par la création de nouvelles infrastructures, d'espaces publics de qualité et de services publics dans le domaine de l'éducation comme les Maisons des Jeunes, longtemps refusés par le régime franquiste pour des raisons de rivalité avec Madrid au plan économique et de surveillance des opinions. Le plan d'aménagement et sa divulgation fonctionnent comme un déclencheur pour la société civile et les associations multiples de Poblenou. L'intellectuel, architecte et géographe Manuel de Solà-Morales y trouve son premier terrain d'étude et de recherche sur la conception et l'importance de la mémoire du tissu urbain dans tout projet d'aménagement, pour ses futurs cours à l'Université Polytechnique de Catalogne (Solà-Morales, 1974).

Les années 1970/1980 voient se fermer les usines les unes après les autres, et le quartier devient une friche urbaine imposante. Seule la fonction d'entrepôts subsiste pendant quelques temps. La Diagonale de Cerdá est coupée par un empiétement de bâtiments disparates et de petites voies de communication rendant illisible le parcellaire initial. Pollution du sol et paysage urbain fragmenté font de Poblenou un territoire en perte de compétitivité et d'attractivité. Quant à la Barcelone olympique de 1992 et ses travaux catalyseurs (Ballester, 2008), ils n'ont pas bénéficié à cette zone, hormis le quartier de la Nova Icaria, le Port Olympique et la réalisation d'une rocade (*ronda littoral*) mais qui ne comporte que deux échangeurs sur le quartier.

Le site et la situation de Poblenou en font un quartier péri-central comportant des aménités valorisantes en cas d'aménagements conséquents (photo 1). Il dispose d'une interface maritime, de la proximité de deux voies de communications majeures en cas de travaux (Ronda et Diagonale), d'un rond point régulateur, et surtout de 198 hectares de surface plane avec un total de 107 îlots disponibles, en partie ou totalement, à la construction. Enfin, les services de l'urbanisme ont acquis une expérience non négligeable avec les Jeux Olympiques : une bonne connaissance de l'*Eixample* (tracé géométrique des îlots) et un savoir faire d'aménageur à partir de très grands projets urbains.

Le seul fait de marcher à travers les ruelles du quartier, de visiter le cimetière de Poblenou, de parler avec les retraités ou de rentrer dans de petits ateliers d'artistes amène à identifier une

---

[2] La mémoire des différents plans d'aménagement et des luttes sociales se retrouvent entre autres sous la forme d'une association historique : les Archives Historiques de Poblenou (Arxiu Històric del Poblenou, Centre Cívic Can Felipa, Carrer Pallars 277, 08005 Barcelona). Pour notre étude, nous mobilisons les caisses d'archives de 1976, 1978 et 2004 comportant les sections : luttes ouvrières, vie locale et nouveau projet 22@, ainsi que la revue de l'association *Icària* complétée par des archives photographiques abondantes (plus de 10 000 clichés).

mémoire collective ouvrière qui a forgé l'histoire du quartier. Les combats contre des projets immobiliers ont représenté des moments d'union et de partage entre habitants, en révolte contre la dictature. À partir du retour à la démocratie, les acteurs de la ville ont dans l'idée de créer une nouvelle Icarie, une ville expérimentale sur le modèle de Cerdá. Mais si les hauts fourneaux, pendant plus d'un siècle, ont permis le consensus par le plein emploi, la période post-industrielle ravive les mémoires et les débats sur ce que doit et peut être le nouveau Poblenou : un complexe d'habitations et un Casino, comme pour le Port Olympique de 1992 (Montaner, 1992) ou une nouvelle utopie socioéconomique ? La spéculation immobilière et les agents du marché songent depuis le milieu des années 1990 à investir cet espace en déshérence de moins de 100 000 habitants et le peuple de Poblenou se pose des questions sur son avenir, celui de ses enfants et parfois celui des artistes ayant réquisitionné des friches. Le quartier attend une action d'aménagement (Tejero et Encinas, 1997). Dans les années 1980, Eduardo Mendoza rappelle dans ses romans l'état d'esprit des décideurs dans le domaine de la stratégie économique :

« *En Catalogne, la tradition, aujourd'hui, est un poids mort. Nous vivons cramponnés au textile et à une petite industrie totalement dépassée. Pour survivre, nous devons nous transformer* » (Mendoza, 2007). Pour la métropole de l'après-JO, comportant plus de 2,5 millions d'habitants avec treize municipalités et une économie urbaine tournée en grande partie sur le « tout tourisme », Poblenou devient un enjeu, celui d'une possible alliance réglementée entre économie créative et attractivité territoriale : l'idée d'une cité digitale comme Nova Icaria commence à germer parmi les décideurs.

Or, les bâtiments en ruine se trouvent occupés par de nouveaux habitants : artistes, travailleurs pauvres, indigents, jeunes, associations et syndicats révolutionnaires. Une prise de possession illégale de nombreuses usines désaffectées se produit dans les années 1990. La mémoire des friches urbaines existe : elle impacte les combats de défense artistiques et, par ricochet, modifie la nature des nouvelles fonctions du quartier créatif. Les artistes deviennent porteurs d'une mémoire par l'intermédiaire de trois facteurs : ils logent dans des locaux rappelant l'histoire industrielle, ouvrière et celle de la production d'objets nobles, en faïence ou en fer forgé ; ils côtoient la population la plus pauvre de Barcelone et connaissent les mêmes problèmes de logement, d'insalubrité et de précarité ; enfin, ils proposent par la création artistique de faire revivre un quartier bien que n'ayant aucune légitimité foncière de par l'illégalité de leur situation[3]. Un esprit du lieu et un génie du lieu se construisent par étape. La centaine de grandes cheminées abandonnées donnent le sentiment que le temps s'est arrêté, mais pas totalement. Un esprit de contestation, de revendication et d'attente se fait jour. L'inertie des pouvoirs publics et la politique du pire des propriétaires désireux d'attendre des jours plus favorables pour la vente de ces terrains (période 1990-2000) expliquent les logiques d'appropriation actuelles d'un espace symbolique et historique de l'histoire de la Catalogne. S'engagent des visions contradictoires de ce que peut et doit être le *nouveau village* du XXIe siècle.

### 1.2. La confrontation de deux logiques d'appropriation de l'espace

Une première logique repose sur la capacité d'une société innovante à proposer un nouveau régime foncier pour Poblenou (1.2.1). Mais les artistes et la population locale de Poblenou restent perplexes face à une telle entreprise et constatent les dégâts, à leurs dépens, d'un urbanisme capitaliste sans âme. Une critique constructive du projet voit le jour, les artistes se faisant géographes et statisticiens de leur disparition programmée (1.2.2).

### 1.2.1. La logique marchande d'un territoire

Les références aux thèses de Richard Florida (2002) et de Michael Porter (1998) sont explicitement revendiquées par les acteurs du projet, notamment en la personne de Miquel Barceló, directeur de la structure 22@, au tout début de sa conceptualisation (Barceló, 2001 ; 2010 : 76).

[3] Travail de terrain, juillet-Août 2007.

Le cercle Digital de l'Institut catalan de technologie dispose de bureaux dans une ancienne usine de Poblenou. Lors d'une conversation avec Ramón Garciá Bragado, alors en charge de l'urbanisme de la ville entre 1999 et 2003, les deux hommes centrent les débats sur la possibilité de faire de Poblenou le Chelsea new-yorkais, mais par le pouvoir de la planification urbaine. Une visite du quartier américain est réalisée, qui donne naissance à une série d'études sur une vingtaine de villes créatives et de quartiers créatifs dans le monde, comme Silicon Alley à New York, Bengalore en Inde, Cambridge en Grande Bretagne ou le parc industriel *Ronneby Software* en Suède. On couple à ces études des visites de terrain à Poblenou, qui présente des similitudes avec le Soho de New York où des collectifs d'artistes se sont installés formant de nouveaux quartiers branchés. Des représentations spatiales sont avancées, comme « le triangle des bars », discothèques, ateliers d'artistes rappelant le New York des années 1970/1980 (Oliva, 2003 : 1-11). La compilation de divers exemples d'espaces de la créativité est contestable car elle prend en considération une multitude de cas qui n'ont ni la même échelle, ni la même histoire, ni la même situation géographique, ni la même culture d'appropriation de l'espace. C'est la technique du *benchmarking*, très répandue en Espagne au tournant des années 2000, notamment dans le domaine de l'aménagement. Quant à la presse grand public, elle se fait l'écho des débats et relate des exemples comme la *Silicon Valley*, ou Palo Alto et ses studios d'artistes. Emerge alors un concept porteur pour la communication et le marketing à l'international : *Poblenou, la nouvelle cité digitale – digital city*.

Dans un second temps de la conception du projet, on imagine un partage de compétences entre le service de l'urbanisme et une entité municipale pour gérer la question de l'emprise foncière entrepreneuriale et celle du locatif et de l'accession à la propriété pour les familles. On promet une croissance économique profitable à tous les concitoyens en intégrant le terme « créativité » dans les discours de communication. Les promoteurs du projet estiment devoir faire face à une concurrence interurbaine européenne qui nécessite d'enclencher un nouveau levier de croissance économique. La création d'un district créatif est au cœur du débat public entre 1998 et 2000 et rentre dans une phase exécutive et de positionnement réglementaire entre 2000 et 2002. Ce projet se lit dans le POS où, au début des années 2000, les urbanistes font passer la zone 22A de Poblenou (Plan Général Métropolitain des années 1954/1976/1980 – A pour zone industrielle) en zone 22@ (PGM révisé à de multiples reprises depuis – @ pour zone de la nouvelle économie). Le A devient @ pour marquer le remplacement de l'emprise au sol par l'industrie de la connaissance et du numérique. L'axiome intellectuel joue énormément dans la conceptualisation de ce que sera le futur Poblenou. Pour la symbolique, il convient de rappeler que, sur le clavier espagnol, le 2 et le @ sont sur la même touche. La tendance est de croire à la création d'une grammaire dite générative, avec un social intégrable dans des réflexions intellectuelles amenant à une formation logico-mathématique de la société, ou modèle systémique en géographie. Une sorte d'automatisme induit par le pouvoir de la technique au profit des hommes. C'est un déterminisme impliquant un scientisme positiviste, souvent à très faible prédictivité puisque c'est sur le marché que repose cette théorie. Les acteurs publics du projet pensent la nouvelle économie urbaine comme une alliance de nature entre la technique et l'économie, avec néanmoins une tendance à la sur-médiatisation et la surestimation du pouvoir de la *techné* au détriment de l'individu qui fait société (Cormerais, 2001).

Figure 1 : confrontation de deux logiques territoriales pour un laboratoire urbain

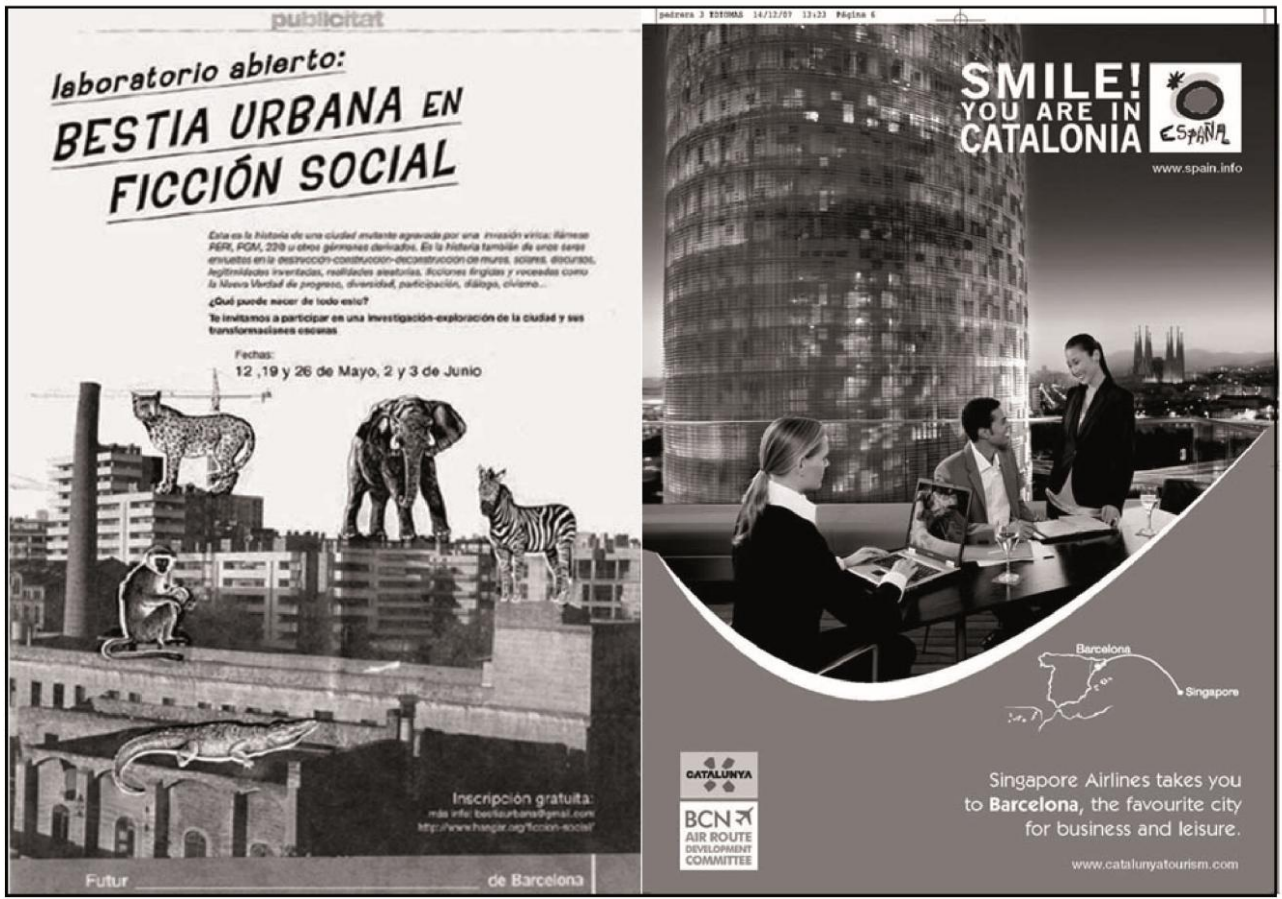

« *Un laboratoire ouvert : bestiaire urbain en fiction sociale. Ceci est l'histoire d'une ville mutante aggravée par une invasion virale : se nommant PERI, PGM, 22@ ou autres germes dérivés. C'est l'histoire aussi d'une série d'entrelacs dans la destruction-construction-déconstruction de murs, d'îlots, de discours, de légitimités inventées, de réalités aléatoires, de fictions, de feintes et de proclamations comme celle de la Nouvelle Vérité du Progrès, diversité, participation, dialogue et civisme… Que peut-il naître de tout ceci ? Nous t'invitons à prendre part à une recherche, une exploration de la ville et de ses transformations obscures les 12, 19, 26 mai, 2 et 3 juin 2007.* »

« *Souriez vous êtes en Catalogne. Singapore Airlines vous amène à Barcelone, une ville privilégiée pour les affaires et les loisirs.* »

Source : Trajet dans Poblenou, parodie d'une publicité municipale, Ramujkic V., Pedrosa T., collectif d'artiste de Hangar, 2006.
Source : Publicité de l'Office du Tourisme de Barcelone, 2007.

Des règles en matières foncières sont formulées : la maîtrise totale du foncier est dévolue à la mairie et aux agents de la société 22@. Pourtant, l'initiative privée est à l'origine de la modélisation du nouveau quartier. Les propriétaires d'îlots doivent rencontrer des promoteurs immobiliers pour concevoir ensemble un projet de rénovation urbaine. Ils négocient dans un rapport donnant/donnant : l'un apporte le terrain, l'autre le savoir-faire du constructeur (avec très souvent des dons d'appartements au propriétaire du terrain). Puis, ces acteurs se rapprochent des services d'urbanisme et du siège du 22@ (entité 100 % publique de la mairie — Arroba s.a.u.). S'engage alors une deuxième négociation sur le plan de masse et la hauteur des bâtiments, notamment sur le coefficient d'occupation du sol à appliquer. Si le projet prévoit de mettre en place une partie de logements sociaux et une implantation d'entreprises reposant sur l'économie innovante, on augmente la densité en autorisant un nombre d'étages plus important. L'aménagement se conçoit à partir du référentiel de l'îlot Cerdá. La ville doit devenir un objet de croissance innovante (Borja, 2004 ; Walliser, 2004) mais c'est une future source de confrontation avec les artistes et la population locale.

### 1.2.2. La logique culturelle et artistique d'un territoire

Face à la réalisation par étape du projet 22@, les artistes se font géographes et statisticiens pour mieux exister, résister et asseoir leurs revendications : ils redessinent la ville, leurs implantations et rationalisent leurs activités pour devenir plus lisibles et par conséquent respectables dans un quartier en chantier. La logique foncière d'un cluster créatif d'une agence municipale affronte la logique de l'artiste se faisant de fait critique social (figure 1).

La disparition des ateliers est un fait. Certes, il est difficile à un moment x de les comptabiliser du fait de la dimension du quartier, de l'insécurité régnant dans certains arrondissements et bâtiments. On se doit de corréler avec prudence plusieurs méthodes de recensement utilisées par la mairie, les artistes, les associations, les partis politiques, les syndicats et la police pour dresser un état des lieux de la présence des ateliers durant la période 1990/2010. De grandes tendances se dégagent et expliquent les stratégies des artistes. En 20 ans (1980/2000), plus de 250 ateliers d'artistes s'installent à Poblenou ; entre 2000 et 2010, plus de 200 ateliers d'exposition doivent fermer leurs portes. En 2006, 12 845 m² sont dédiés à la production artistique. Dans la seule année 2007, 18 espaces d'exposition comportant plus d'une salle ferment, soit 6783 m² d'exposition en moins, d'où la disparition de 200 artistes. On compte en 2008/2009, une cinquantaine d'ateliers réguliers possédant des locaux d'expositions, exposant et participant à la vie culturelle du quartier. Fin 2010-début 2011, il ne reste plus que 7 grands ateliers, dont Hangar, le plus reconnu de tous au plan national et international. Des cartes et flyers mettent en parallèle l'avancée des travaux du 22@ et la disparition des ateliers. Une réflexion s'engage sur les moyens de résister ou de composer avec l'agence municipale d'aménagement.

### 1.3. L'usine Can Ricart, révélateur d'un malaise urbain ?

Une mobilisation populaire va devenir l'emblème de la révolte contre le système mis en place par les agents du 22@ sur quatre îlots concentrant la majorité des ateliers fin 2005. La diversité et l'originalité des propositions des artistes pour sauver l'un des fleurons du patrimoine industriel catalan, Can Ricart, instaure une bataille urbaine populaire et médiatique trouvant un certain relais dans la population du quartier. Il s'agit d'un moment de cristallisation et de crispation, une lutte urbaine comme dans la banlieue de Madrid où Manuel Castells (1983) rappelle trois facteurs explicatifs : le décalage entre l'offre et la demande de service public de la part des acteurs et de la population, la défense d'une identité culturelle d'un groupe lié à la défense d'un territoire, et surtout qui contrôle ou croit vraiment contrôler le territoire.

---

[4] www.salvemcanricart.org, site de la plateforme de contestation, consulté le 01/07/08.

La lutte pour la sauvegarde et le sauvetage de Can Ricart est un exemple de transition territoriale par une lutte urbaine amenant en partie à la redéfinition du projet urbain (Martí, 2008). Un infléchissement de la stratégie foncière se concrétise à la suite de longues manifestations dont les acteurs en présence sont : un propriétaire impatient de vendre les parcelles de son îlot, des artistes/squatteurs en quête d'une sécurité de logement et d'ateliers de travail, une société publique 22@ brandissant son cahier des charges, une mairie incapable de prendre des décisions cohérentes et une opinion publique regardant les péripéties juridiques et épisodes violents avec circonspection. Une lutte citoyenne, intellectuelle et artistique s'engage pour Can Ricart tout en espérant un changement d'orientation des plans du 22@ (Broggi, 2007). Le combat ne naît pas du hasard des convulsions d'un quartier.

Des gratte-ciels apparaissent au nord et sud de Poblenou avec deux secteurs clefs de l'opération : les parcelles limitrophes de la grande avenue Diagonal où se trouve Can Ricart et le quartier du Forum 2004 inauguré lors de la grande fête mondiale basée sur la culture, fortement critiquée lors de sa tenue (Ballester, 2013a). Les artistes craignent l'arrivée inopinée des *mossos d'esguardas* (CRS régionaux) pour les déloger. Certains négociants et propriétaires d'ateliers artisanaux participent également à la lutte car ils craignent de perdre des entrepôts, ateliers de travail parfois informels mais bon marché et bien situés. On note, dans le mouvement contestataire, des historiens luttant pour la mémoire des classes ouvrières en compagnie d'intellectuels ou universitaires

Figure 2 : Des bicyclettes populaires pour sauver Can Ricart et présentation du site web

Source : Affiches, flyers du Collectif Sauvons Can Ricart ; 2006, 2007.

barcelonais de la nouvelle économie déterminés à prouver la possibilité de faire cohabiter artistes et industries de la connaissance. Le plus connu de ces intellectuels est Josep Maria Montaner, historien, essayiste sur la ville de Barcelone (Montaner et Muxí, 2006).

Une véritable montée en puissance des actions collectives se produit avec des articles militants sur Can Ricart dans des revues comme l'*Avenç*. Des maquettes d'architecture sont présentées sur le site même, avec des possibilités de sauvetage total ou partiel. Elles constituent des contre-projets d'architecture et d'aménagement fort bien documentés par des séries d'expositions publiques ou des manifestes dans des revues en ligne (Groupe du patrimoine industriel, du Forum de la Ribera du Besós). En 2005, la procédure de réquisition des locaux et des expulsions s'accélère tout autour de l'usine. Federico Ricart, propriétaire et héritier du passé industriel des lieux, est fort mécontent d'un squat devenu légitime et emblème de la lutte pour une partie de la population. Il demande l'expulsion des artistes. « *Salvem Can Ricart* » — Sauvons Can Ricart — est créé[4]. Une assemblée est organisée sous l'impulsion de l'association Hangar comme fer de lance de la contestation puisqu'elle est directement menacée. Un mouvement de revendication voyante et quotidienne s'organise par des marches et des manifestations qui se tiennent régulièrement place Saint Jaume, en centre-ville, pour sensibiliser toute la population (figure 2).

Douze mois d'occupation de l'usine, suivis d'une expulsion mouvementée par la police et de la fermeture du squat, obligent la mairie à proposer à son tour un Plan Stratégique de la Culture via l'Institut Culturel de Barcelone. La visibilité des artistes et la mise en scène d'une expulsion contraignent à restituer une partie des locaux jouxtant l'îlot principal, devenu insalubre, à Hangar, association d'artistes ayant négocié avec la mairie et devenant de fait un interlocuteur privilégié de cette dernière[3]. L'année 2006 permet par la législation de caractériser Can Ricart

Figure 3 : Le nouveau Can Ricart à l'horizon 2014 comme bien d'intérêt culturel patrimonial en Catalogne. Or, fin 2006, un incendie ravage la zone centrale de l'usine désaffectée. La partie sauvée des flammes est alors concédée à la mairie et à une association pour un futur projet ambitieux au plan artistique ; une autre partie est définitivement démolie et la dernière partie revient de droit à Federico Ricart. Le propriétaire estime qu'il a subi une expropriation frustrante mais il n'a jamais compris l'aspect symbolique de la bâtisse, rompant avec le passé d'industriel social de ses aïeuls.

Paradoxalement, après cette catharsis impliquant toutes les couches de la société de Poblenou, une solution est trouvée justement par l'affectation d'un COS de plus de 2,4 sur la parcelle. Les trois parties en présence utilisent les mécanismes du 22@ pour trouver un compromis grâce à l'échange foncier et à l'autorisation de construire des immeubles de rapport d'une quinzaine d'étages. En 2009, la mairie élabore une programmation ambitieuse de reconstruction et de rénovation par l'implantation de services publics sur le site de l'usine. Une maison des langues (figure 3) financée par la mairie et la région, en lien avec les revendications autonomistes/indépendantistes catalanes et l'année des langues de l'ONU en 2009, est proposée à côté de lofts, de bureaux et d'un espace culturel contemporain (non encore identifiable, un musée, une zone mixte de recherche et d'aide à la créativité). Telle est la destinée future de ce lieu devenu emblématique et marqueur d'une transition territoriale à la suite d'une revendication artistique. Quant à l'usine, elle devient objet d'étude dans les écoles, centres d'architecture ou culturel et retrouve même son aspect d'antan sur les timbres officiels espagnols (figure 3).

Le statut marginal des artistes devient une position de respectabilité et de responsabilité auprès de toute la société. Le changement social est réel. Non contents de résider sur un territoire stratégique, ils impulsent à leur tour des stratégies de survie devenant tactique pour créer un espace de dialogue et de compromis pour la réalisation d'un district innovant faisant explicitement référence à leurs actions. Reste à entrevoir sur le long terme les premiers résultats d'une alliance au départ contraire entre artistes et cluster innovant, et devenue source de croissance et d'expérience depuis. L'artiste a révélé la valeur du quartier au reste de la population par un jeu complexe de retour sur une mémoire collective parfois parcellaire, anachronique et restituant des bribes de vérité et de mystifications. Peut-il rentrer dans la politique officielle de créativité d'un cluster ?

---

[3] http://barcelonacultura.bcn.cat/en/culture-institute, site de l'Institut Culturel de Barcelone, consulté le 01/07/2010.

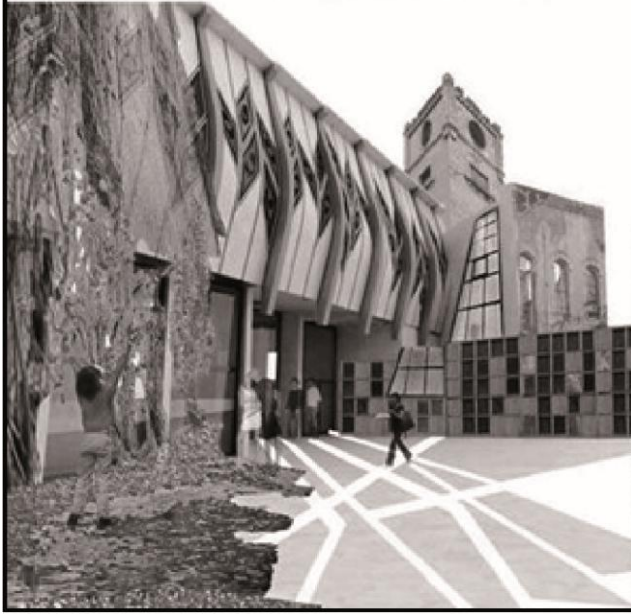
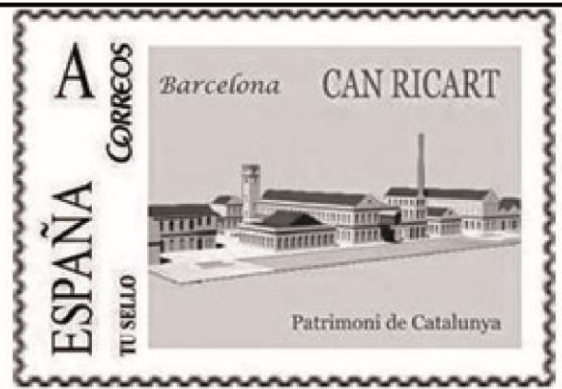
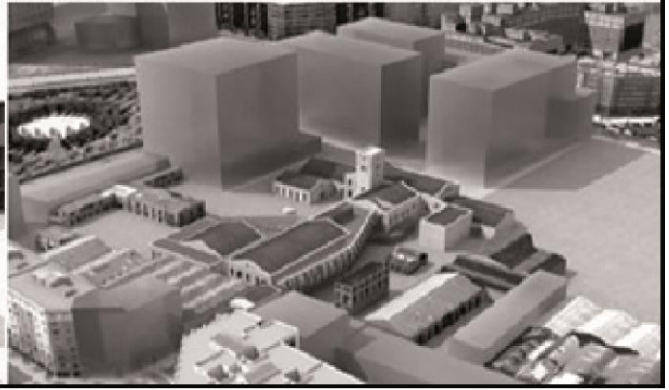

Source : Mairie de Barcelone et Poste Espagnole, 2010.

## 2. LE DISTRICT 22@ : VERS LA PRODUCTION D'UN ESPACE DE COMPROMIS ?

L'enjeu patrimonial est un révélateur des nouvelles interactions entre artistes et cluster économique, surtout à Barcelone où les règles de construction ne sont en rien des freins à l'inventivité ou à l'originalité, comme elles peuvent l'être en France avec les ZPPAUP (2.1). Au plan physique, matériel et architectonique, des cohabitations sont donc possibles, mais également au plan intellectuel où des passerelles sont proposées entre artistes et entreprises innovantes (2.2). Il reste cependant à comprendre, par l'intermédiaire des derniers chiffres de l'agence 22@ et du ressenti de la population locale, les dégâts d'une société ayant basé sa croissance sur l'immobilier de rapport (2.3).

### 2.1. Le patrimoine industriel comme reflet d'une nouvelle identité territoriale

Les rapports d'activités de 2000/2005 de l'agence 22@ minorent les références à la protection du patrimoine. Il faut attendre les mémoires d'activités de 2006/2010 pour voir apparaître une section « respect et protection du patrimoine industriel » : « *un Plan Spécial de Protection du Patrimoine Industriel de Poblenou qui prévoit la conservation d'un total de 114 éléments (46 qui sont déjà catalogués et 68 sont en passe d'être inclus) qui contribuera à garantir la préservation de ce legs d'intérêt historique et culturel* » (Ayuntamiento de Barcelona, 2010 : 11). Depuis Can Ricart, et de manière plus autorisée, l'agence réagit dans une volonté de créer une identité à part entière du quartier comme une sorte de réconciliation entre tous ses habitants. Un diagnostic précis accompagne ces marchandages. Dès qu'ils sont approchés par les propriétaires et les promoteurs immobiliers, 22@ et les services de l'urbanisme de la municipalité regardent les enjeux patrimoniaux. Il leur incombe de garder la spécificité de Poblenou dans un mélange entre anciens et nouveaux bâtiments. Des parties d'anciennes usines peuvent côtoyer à seulement quelques mètres des immeubles contemporains imposants, pour le néo tertiaire. Les exemples de l'université *Pompeu Fabre* et sa place Gutenberg avec sa grande cheminée (photo 2), de l'université de Catalogne et ses annexes d'anciens entrepôts et de l'îlot *Can Jaumandreu* sont probants. Cette dernière opération a réhabilité l'ossature extérieure de la filature lainière *El Vapor de la Llana* édifiée en 1870 et comprend depuis un centre moderne de bureaux.

Si on définit l'innovation sociétale comme un système flexible faisant de la créativité une source de souplesse dans un environnement en construction dit durable (Phills *et al.*, 2008), on peut émettre l'hypothèse que cette innovation sociétale propose des opérations de redéfinition d'un projet urbain face aux revendications sociales, comme dans l'exemple de Can Ricart. Pour Barcelone, en quête de compétitivité et d'originalité, il est nécessaire de recréer un esprit du lieu et de l'entretenir en associant créativité et héritage urbain. À une architecture de signature de type Guggenheim à Bilbao, on ajoute une obligation d'inclure dans le dessin les éléments du passé de Poblenou. L'architecture entretient le sentiment de créativité qui permet « *la capacité à diffuser le marquage de l'espace à travers divers systèmes de repères symboliques devenant une préoccupation fondamentale de l'aménagement de l'espace* » (Castells, 2009). La flexibilité du projet le rend cohérent en partie. Les premiers résultats montrent un paysage urbain tout à fait soutenable grâce au pragmatisme catalan et à la volonté de dépasser les frontières établies en ce domaine.

Source : P. Ballester, 2008.

Photo 2 : Université *Pompeu Fabre*, place Gutenberg, 2008

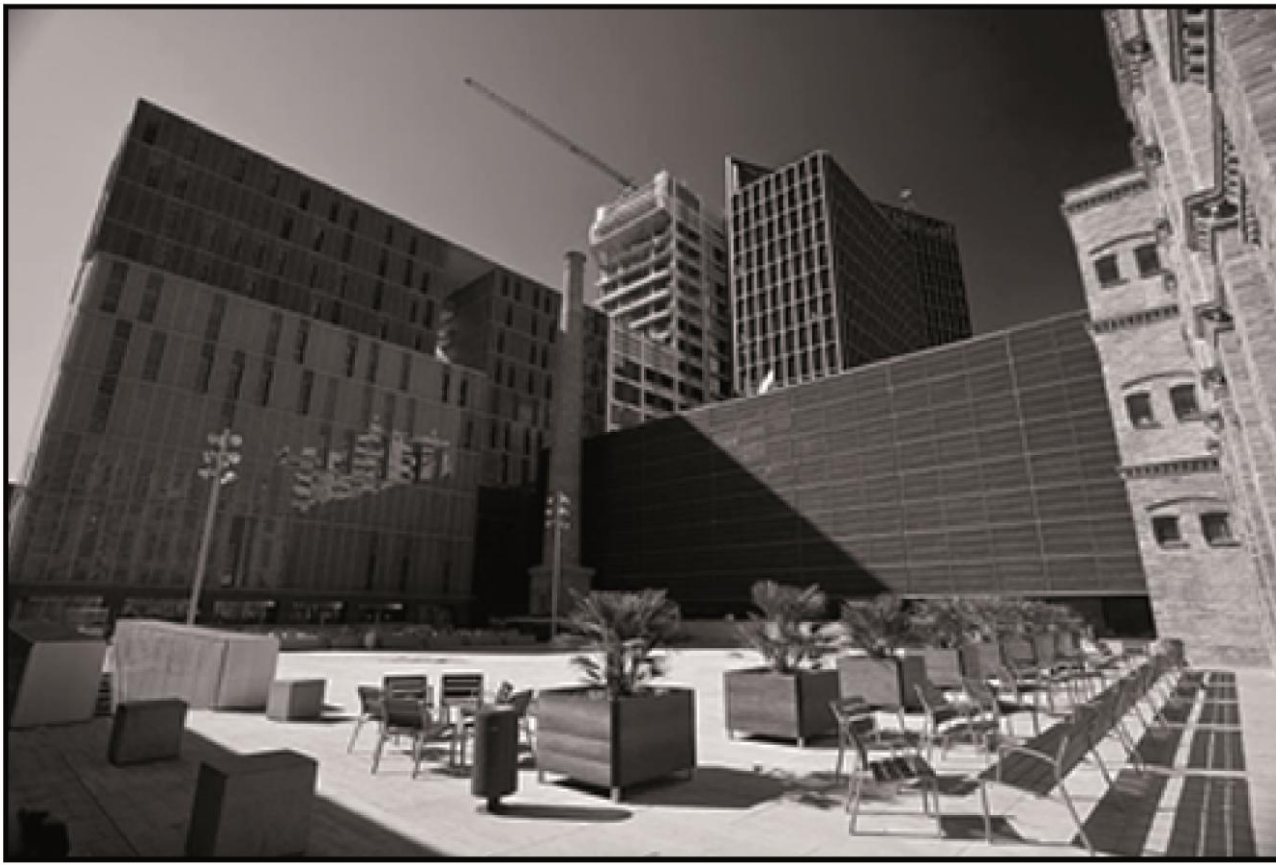

### 2.2. Les tentatives de collaboration entre artistes, entreprises et population locale

Reste à évaluer les résultats, pour les artistes en présence sur Poblenou, d'un pragmatisme à la fois artistique et de subsistance. Pedro Soler, directeur de Hangar au plus fort de la lutte, propose de ne pas être dupe des mutations territoriales d'une part, et d'investir, d'autre part, le champ des nouvelles technologies en lien avec la création à Poblenou : « *La créativité est un terme à la mode ; pour sauver le capitalisme, les artistes peuvent offrir leurs talents tant aux industries culturelles que pour les expositions, les fêtes… dans lesquels les jeunes et les intellectuels veulent vivre et se développer. Hangar ne peut et ne doit pas éviter ce processus d'intégration. […] L'intégration de l'art et des artistes dans le nouveau paradigme de cette société doit être accompagné d'une*

*redistribution correcte et juste des ressources. De plus en plus, nous devons prendre en compte ces opportunités* » (Soler, 2009 : 7-10).

L'insertion dans la sphère économique des activités créatives et éducatives des artistes de Poblenou se fait timidement. Un rapprochement est à souligner entre les activités créatives artistiques et les grands groupes du district. L'exemple le plus parlant provient encore une fois de Hangar, l'une des rares entités à pouvoir nouer des liens avec ces firmes sur la longue durée. Le Parc de recherche Biomédical de Barcelone propose une collaboration éducative et artistique par le biais d'expositions multimédia interactives sur des critères didactiques pour les lycées et collégiens barcelonais ainsi que l'utilisation de techniques nouvelles de micro-rayons lumineux permettant l'inscription sur les murs de phrases, de résumés de parcours éducatif ; l'utilisation de la musique électronique rappelle quant à elle la présence des sons quotidiens d'un laboratoire[6]. Il en est de même avec le *Summer Lab Camp* : le principe de la manifestation réside dans la diffusion et l'exposition de travaux numériques, vidéos et internet en lien avec les écoles de design et de communication, qui regardent avec curiosité et perplexité cette insertion créative dans leur sphère éducative réglementée. Il reste cependant une barrière infranchissable : il n'existe pas véritablement de correspondance régulière, suivie et pécuniaire avec le système éducatif et les écoles de Design ou de Télévision de Barcelone (cluster Media-Tic), lesquels font appel uniquement à des professionnels. L'articulation entre activité créatrice et industrie créative tend à se faire au profit des majors, tout en évacuant le noyau originel des artistes alternatifs au profit des institutions ou associations reconnues, comme Hangar. Les tentatives d'association avec les musées, réseaux scolaires, fêtes électroniques et rencontres sur le terrain pondèrent les processus, sans toutefois totalement les inverser.

On note l'évolution des propos de chaque partie en présence. L'étrange rencontre des deux champs lexicaux, marchands et artistiques, est à souligner car il montre une alliance de fait. L'échelle du territoire d'action, de représentation et d'exposition se révise dans sa globalité compte tenu que le quartier devient désormais un quartier international, avec l'arrivée de nouveaux modes de transport comme l'AVE, train à grande vitesse espagnol (figure 4). Les artistes et dirigeants de Hangar parlent « *d'échelles de travail superposées : le quartier, la ville, la Catalogne, l'Espagne, l'Europe, l'Amérique du Sud et le monde. Elles doivent se nourrir mutuellement* » (Soler, 2006 : 7). Sur les quinze ans à venir, les artistes estiment possible d'entretenir des liens réguliers avec des correspondants lyonnais. La politique culturelle est simple : plus de coopération, plus d'ouverture pour plus de mécénats privés ou publics de grands groupes, comme il est coutume en Espagne. Les perspectives financières et la géographie du site et de la situation de Hangar par rapport à la future gare de Sagrera toute proche [6] Observations de terrain : manifestation d'octobre 2010. devient une légitimation de l'action culturelle par réseaux concentriques : ils utilisent les mêmes arguments que le 22@.

Des efforts sont entrepris pour proposer une image positive des artistes qui demandent à participer aux fêtes locales du quartier. Compagnies artistiques et troupes de rue intègrent difficilement et peu à peu la *Feste de Maig, la Festa Major*. Ils participent à des moments de culture comme *Can Felipa, Escena Poblenou*. De même, avec l'aide de fondations, comme la Banque Sabadell, ils créent le Département des Nouveaux Talents ainsi que le *Hybrid Playground de Lalalab ou le wip*, manifestations basées sur la création numérique, qui sont à la fois en concurrence et en lien (peut être en coopération prochainement) avec le festival officiel de la ville, le BVS (figure 5).

La question de la professionnalisation et de l'intégration des artistes dans des secteurs marchands reste un handicap pour ces ateliers, mais peut aussi être une future ouverture. Leur fragilité repose sur celle du financement et sur la réception des artistes par les habitants, sans parler des réductions de budget des mécènes face à la crise économique. L'institutionnalisation de la créativité produit essentiellement des frustrations pour ceux qui en sont bannis, car elle propose des procédures de bourse et un montage de financement de projets culturels qui s'apparentent aux codes régissant les relations entrepreneuriales et les guerres commerciales entre firmes multinationales.

La créativité, la compétitivité et la compétition structurent désormais les rapports entre les artistes et au sein même du cluster. Il est étrange de constater que les ateliers fonctionnent comme des entreprises, en répondant à des appels d'offre ou en subissant des audits. Néanmoins, si l'on étend le champ de nos investigations à tous les citoyens, on peut alors parler d'échanges citoyens dans le quartier. Des exemples, comme *la mémoire de nos ancêtres*, proposent par le biais du numérique de faire se rencontrer de jeunes artistes, des créateurs et des personnes âgées, pour raconter et exposer l'héritage symbolique et associatif du quartier.

Figure 4 : Les transports, facteurs de coopération artistique. La future gare Sagrera et le réseau de l'AVE

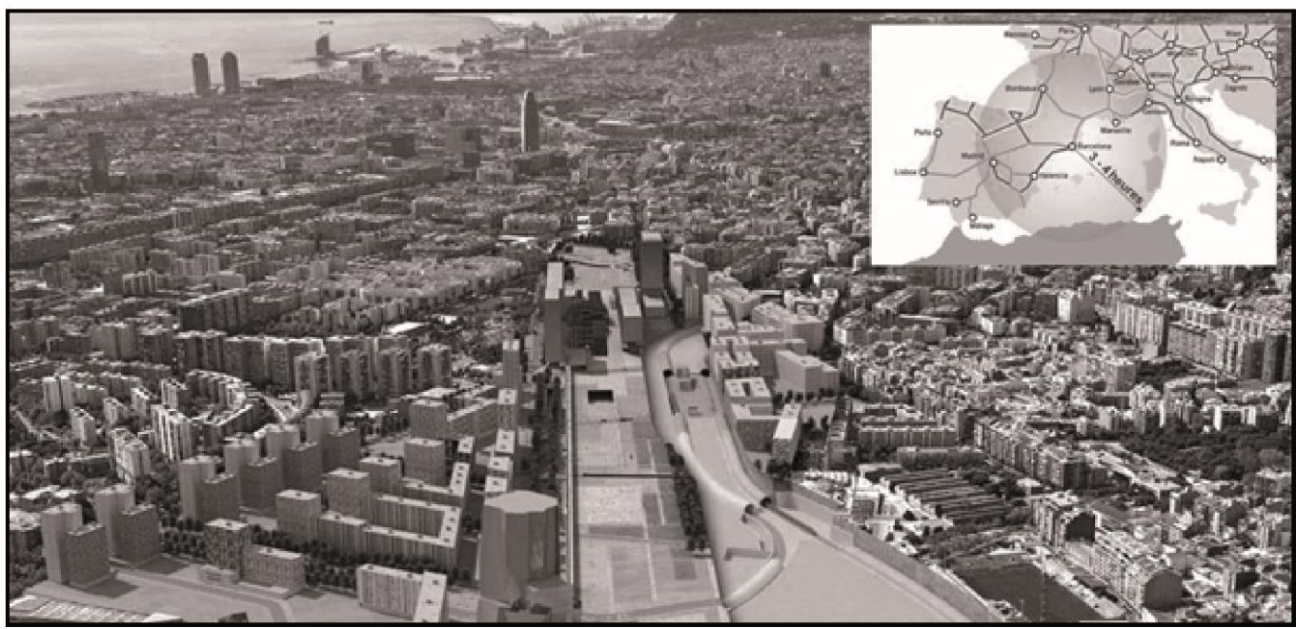

Source : avec l'aimable autorisation de Sagnera projet, 2010.

Figure 5 : Le BVS et le Wip : juxtaposition ou coopération ?

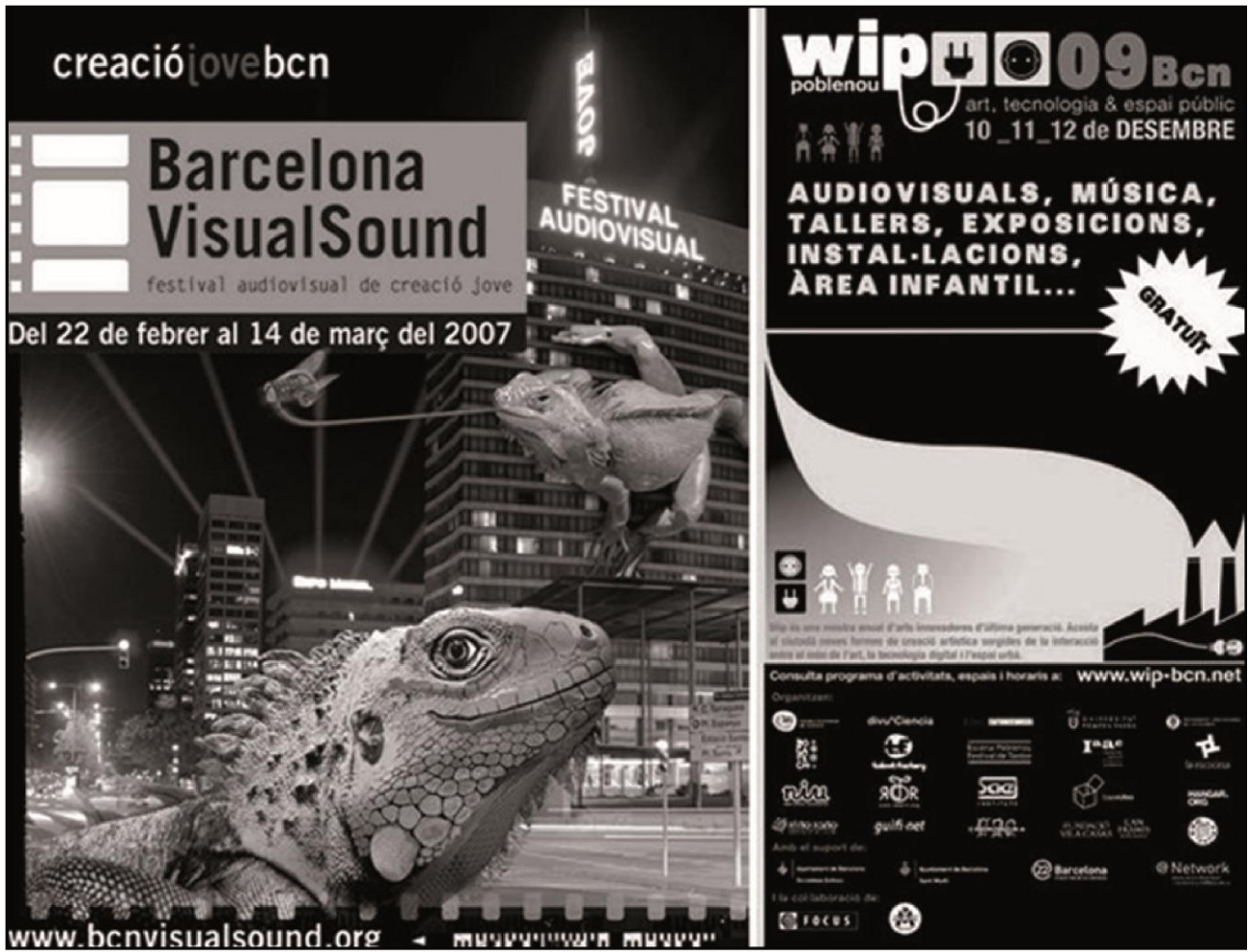

Source : avec l'aimable autorisation de BCN et Hangar (2007, 2009).

De même, avec le *22@CreaTalent*, les firmes, et uniquement elles, proposent une initiation aux outils informatiques aux enfants du primaire. Enfin, le *computer recycling* permet aux entreprises de 22@ de donner des ordinateurs à la population et à quelques artistes en manque de budget. Lier technologie, mémoire, aide associative et talent artistique et d'écriture correspond à un cahier des charges ambitieux qui implique connaissance et respect mutuel de l'apport de chacun dans le nouveau Poblenou.

**2.3. Dix années de régénération urbaine : entre crise morale et raisons d'espére**r

Des résultats probants montrent le décollage du cluster et l'adaptation des artistes à un mode opératoire reposant sur l'économie de marché tout en agissant dans le sens de la redéfinition du projet 22@ (Arroba. S.a.u., 2010). Mais ces résultats restent fragiles du fait du contexte économique espagnol et européen. Concrètement, en dix ans d'action, 1502 entreprises se sont implantées à Poblenou, ainsi que 12 centres de recherche et de développement. Douze établissements supérieurs se sont implantés dont 5 universités, 3 écoles de commerces, 2 écoles de design et 2 écoles de communication-marketing (Casellas et PallaresBarbella, 2009)[4]. On notera qu'il s'agit non plus d'un territoire pauvre, mais riche, car en dix ans de mutation le volume des transactions s'élève à plus de 6 milliards d'euros. Or, quelle est la part d'emplois des habitants de Poblenou dans ces nouvelles industries ? Elle est minime : moins de 5 % du total des nouveaux postes, avec souvent des tâches subalternes et répétitives, donc en rien créatives. De plus, les classes dites créatives sont plus des agents de service du néo-tertiaire que des créatifs. Il faut attendre le prochain recensement officiel catalan pour connaître leurs lieux de résidence : le quartier de Poblenou ou la banlieue du grand Barcelone ? Au plan métahistorique, une réussite certaine du projet est à entrevoir : en 2008, le Poblenou a totalisé davantage de créations d'emplois qu'à l'époque de l'antique zone industrielle, il y a quarante ans au plus fort de son activité, soit 40 000 emplois actuellement contre 35 000 au moment du sommet industriel des années 1955/1965 (Sodupe, 2007). Restent cependant des tensions urbaines qui montrent le manque de concertation ou d'obligation dans le partage foncier. Les publicités de promoteurs vantent un quartier à la mode mais nient le fait que Poblenou reste un vaste chantier, une zone de non droit pour habitants squatteurs, un lieu de boites de nuit bruyantes, et surtout une zone de politique innovante et un modèle pour une équité sociale qui n'a réalisé que 7 % des logements sociaux prévus. Car les artistes sont aussi les premiers à revendiquer des logements sociaux en péri-centre de Barcelone. 576 logements sociaux sur les 40 000 attendus à horizon 2020 ont été réalisés. L'Espagne, en pleine crise hypothécaire, a stoppé le rythme des constructions à Poblenou mais aussi leur vente : seulement 8556 habitations ont été construites entre 2009 et 2011, et sur les 42 619 logements construits entre 2000 et 2010, 5 504 n'ont toujours pas trouvé acquéreur en 2011. Le recul des ateliers collectifs d'artistes est une réalité depuis plus de dix ans. Les agences immobilières proposent des offres de locations d'ateliers au prix du marché. Les prix des lofts et appartements de standing proches des anciennes usines occupés sont à 7000 euros le m² comme base de négociation (Narerro, 2003). Mais surtout, l'ensemble de la population de Poblenou doit affronter un chômage de masse et un ralentissement conséquent de l'activité économique, ce qui infléchit le rythme des implantations dans le cluster et le mécénat envers les ateliers d'artistes. Enfin, l'héritage industriel et le manque d'industries de pointe en Espagne expliquent le décollage timide du cluster TIC Médical (Lazzeretti, Boix et Capone, 2008).

**CONCLUSION**

L'expérience Poblenou du début du XXIe siècle est une remise en cause des frontières de la gestion urbanistique et culturelle d'un territoire hautement spéculatif par l'action symbolique des artistes et de la population locale de Poblenou. Ainsi, se confectionne une société du dialogue qui se base sur l'interdépendance et la réciprocité des perspectives. Nous assistons à la création d'un espace urbain de compromis et d'échange, régulier mais timide dans sa forme et sa façon de concevoir *l'autre* comme une chance et non un danger. Les artistes sont les portevoix et initiateurs de changement dans la gouvernance foncière et patrimoniale, et dans la gestion culturelle du district. Un quartier créatif laboratoire procure de nouvelles relations entre ses composantes. Pourtant, un malaise profond subsiste de par les regards ambigus portés par les institutionnels catalans (Grésillon, 2008). Trois facteurs consubstantiels du nouveau Poblenou sont prégnants :

- Une mémoire collective résiduelle et parfois réinventée s'est forgée sur plus d'un siècle, avec une valorisation de la mémoire des anciennes luttes pour justifier les nouvelles luttes malgré les anachronismes, redirection, embellissement et intermittence que cela engendre (Lavabre, 2001). Cette lutte doit engendrer de nouveaux rapports de force et de nouvelles instances de vie entre anciens et nouveaux habitants. Can Ricart en fut un laboratoire.

---

[4] Sources : Joves de Poblenou, site des jeunes de Poblenou (Syndicat), www.jovespoblenou.cat consulté le 01/05/11. Chiffres d'Arroba. S.a.u., 2010 et 2011 ; il faut être prudent avec les chiffres du 22@ car les données sont avancées à grand renfort de communiqués de presse.

- Le choix entre architecture contemporaine et protection patrimoniale s'est résolu dans une alliance entre les deux, dont l'îlot de Cerdá est le moteur. Le quartier créatif s'en retrouve embelli et donne une image positive à l'international.

- L'artiste est un nouveau créateur de figure urbaine complexe dans sa capacité à profiter de situations conflictuelles pour exister et prendre l'ascendance sur d'autres groupes inorganisés.

L'exemple du collectif Hangar montre qu'à partir de situations conflictuelles, une nouvelle légitimité se construit. Enfin, vie et mort d'une ville européenne vont de pair avec la gentrification (Jacobs, 1961) : celle-ci s'observe bien mais relève plus d'une particularité locale que de la ville dite créative (Bourdin, 2008). Le plus inquiétant est certainement le nombre de locaux ou logements vides dans un cluster qui est finalement plus d'ordre marketing que créatif (Saez, 2010). Les *Living Lab* de type scandinave ouverts au grand public sur la thématique des nouvelles technologies sont peut-être une solution de rapprochement. La société marchande espagnole du XXIe siècle produit un espace néo-capitaliste dominateur et fragmenté par une frénésie immobilière irréfléchie. L'unité entre l'espace physique, l'espace mental et l'espace social n'était que théorique et imparfaite dès les origines de sa conceptualisation. Si chaque société doit produire son espace, ce fut ici celui de la spéculation dont la véritable valeur resurgit dans les épreuves et les luttes urbaines, dans une production par analogie et parfois par une demande sociale cristallisant les mécontentements, comme pour Can Ricart. Un paysage urbain contemporain de forme hybride se construit par étapes ; un sentiment de beauté retrouvé et de fierté locale se remarque comme les références artistiques à ces anciennes blouses ouvrières dans des tableaux contemporains ou les tags sur des murs délabrés : « *ces lambeaux d'habillement, que ce peuple artiste drape encore avec art* », comme disait Madame de Staël.

**Bibliographie**